\documentclass[doublecol]{epl2}
\usepackage{amsmath}
\usepackage{amssymb}
\usepackage{graphicx}% Include figure files
\usepackage{dcolumn}% Align table columns on decimal point
\usepackage{bm}% bold math
\def\be{\begin{eqnarray}}
\def\ee{\end{eqnarray}}
\def\ba{\begin{array}}
\def\ea{\end{array}}

\begin{document}
\title{Evanescent incompressible strips as origin of the observed Hall resistance overshoot}
\author{A. Siddiki,\inst{1,2} S. Erden Gulebaglan,\inst{3} N. Boz Yurdasan,\inst{3} G. Bilgec,\inst{3} A. Yildiz\inst{3} and I. Sokmen\inst{3}}
\shortauthor{A. Siddiki \etal}
\institute{
  \inst{1} Istanbul University, Physics Department, Faculty of Sciences, 34134-Vezneciler-Istanbul, Turkey\\
  \inst{2} Harvard University, Physics Department, 02138 Cambridge MA, USA\\
  \inst{3} Dokuz Eylul University, Physics Department, Tinaztepe Campus, 35100 Izmir, Turkey }
%\address[l2]{Trakya University, Physics Department, Faculty of Arts and Sciences, 28100 Edirne, Turkey}
\abstract{In this work we provide a systematic explanation to the unusual non-monotonic behavior of the Hall resistance observed at two-dimensional electron systems.
We use a semi-analytical model based on the
interaction theory of the integer quantized Hall effect to
investigate the existence of the anomalous, \emph{i.e.} overshoot, Hall resistance $R_{H}$. The observation of the overshoot resistance at low magnetic field edge of the plateaus is elucidated by means of overlapping evanescent incompressible strips, formed due to strong magnetic fields and interactions. Utilizing a self-consistent numerical scheme we also show that, if the magnetic field is decreased the $R_{H}$ decreases to its expected value. The effects of the sample width, temperature, disorder strength and magnetic field on the
overshoot peaks are investigated in detail. Based on our findings, we predict a controllable procedure to manipulate the maxima of the peaks, which can be tested experimentally. Our model does not depend on specific and intrinsic properties of the material, provided that a single particle gap exists.}
%\begin{keyword}
%Quantum Hall effect\sep overshoot \sep disorder \sep resistance \sep temperature
%\PACS 73.20.Dx, 73.40.Hm, 73.50.-h, 73.61,-r
%\end{keyword}
%\end{frontmatter}
%
%Sinem Erden Gulebaglan, Dokuz Eylul University, Physics Department, Tinaztepe Campus, 35100 Izmir, Turkey\\
%Fax: +90 232 4128888, Email: sinem.gulebaglan@deu.edu.tr\\

\maketitle
\section{Introduction}
One of the most commonly used material characterization method is to measure the resistance of the sample. The current-voltage (\emph{I-V}) measurements provide information about the scattering mechanisms, \emph{i.e.} the properties of the impurities. The resistivity of a material is usually determined by its intrinsic properties, however, these properties might also depend on the external parameters such as temperature $T$, external magnetic field $B$ \emph{etc}. Moreover, if the sample is subject to a perpendicular $B$ field and the transverse resistivity (namely the Hall resistivity) is measured, one can also determine the type and the number density of the charge carriers. The Hall resistivity is linear in $B$ for a typical three-dimensional materials, which is drastically altered at two-dimensional systems to a stepwise behavior. This phenomenon is known as the integer quantized Hall effect (IQHE)~\cite{vKlitzing80:494}. In the early days of the IQHE, the quantized plateaus are attributed to the intrinsic properties of the system together with the gauge invariance principle~\cite{Laughlin81,Thouless82:405}. This picture is named as the localization theory~\cite{Kramer03:172} and is still widely used to explain the IQHE at low-mobility ($< 10^6$ V/cm.s) and large ($> 20~\mu$m) samples. Interestingly, the transitions between the quantized Hall plateaus are commonly linear in $B$ similar to the classical Hall effect, also in 2D. However, non-monotonic anomalous peaks at the low $B$ side of the plateaus are also reported in many different materials and are discussed in a fairly varying contexts~\cite{RIC1992,Ram1998,Gri2000,Shl2005,Shl2006}. The non-monotonic increase of the Hall resistance is known as the overshoot, and is usually attributed to impurity effects, similar to IQHE. Recently, resistance overshoot is experimentally investigated in relatively narrow samples and the results are discussed in the context of interaction induced incompressible strips in a phenomenological manner~\cite{sailer:10}. If a two-dimensional electron system (2DES) is subject to a perpendicular $B$ field, by the virtue of Landau quantization and direct Coulomb interactions, the electronic system is composed of compressible and incompressible strips. The Fermi energy is pinned to one of the Landau levels at the compressible regions and falls in between quantized levels at the incompressible regions~\cite{Chang90:871,Chklovskii92:4026}. Note that the local filling factor $\nu(x)=2\pi l^2 n_{\rm el}(x)$ is an integer at the incompressible strips, where $l=\sqrt{\hbar/eB}$ is the magnetic length and $n_{\rm el}(x)$ is the local electron number density. These co-existing strips change the intrinsic properties (\emph{e.g.} screening, conductivity) of the electronic system considerably~\cite{Lier94:7757}. The compressible regions behave like a metal, due to high density of states (locally), whereas the incompressible can be considered as insulators. The existence and transport properties of these regions strongly depend on the temperature and $B$ mainly, among other system parameters~\cite{siddiki2004}.

Here, we present our results that provide a self-contained calculation scheme to explain the observed resistance overshoot within the interaction picture of the IQHE~\cite{siddiki2004,Guven03:115327}. First we discuss the formation of the compressible/incompressible strips depending on the sample properties giving an analytical description together with a simplified transport calculation. We show that, under certain conditions two or more (evanescent) incompressible strips assuming different filling factors can co-exist and contribute to the current, hence Hall resistance. In the next step, we calculate Hall resistances within a local version of the Ohm's law and numerically investigate the dependencies of the overshoot on temperature, disorder and sample size. At a final section, equipped with the outcomes of interaction based model, we predict that the resistance overshoot can be manipulated by changing the edge profile of the system.
\section{The semi-classical model}
In this section we investigate the formation and deformation of the incompressible strips within the frame work of Chklovskii \emph{et.al.}~\cite{Chklovskii92:4026} and Siddiki \emph{et.al}~\cite{siddiki2004}, respectively. The earlier work is constructed on the electrostatic equilibrium condition of the 2DES in the presence of an external $B$ field. It is assumed that the 2DES resides on the $z=0$ plane, confined by an electrostatic potential due to a homogeneous donor layer with a constant density $n_0$ and the electrons are depleted from the edges by an amount of $l_d$, by the virtue of metallic in-plane gates. Translational invariance is imposed in the $y$ direction. Then, the solution of the Poisson equation satisfying the above discussed boundary conditions together with the analytic continuation yields an electron density distribution of the form~\cite{Chklovskii92:4026}
\be n_{\rm el}(x)=\bigg(\frac{x-l_d}{x+l_d}\bigg)^{1/2}n_0.\ee
It can be shown that, a dipolar strip (\emph{i.e.} incompressible strip) should form in the vicinity of $x_k$, which is determined by the condition $k=2\pi l^2 n_{\rm el}(x_k)$, $k$ being an integer. The incompressible strip is electrostatically unstable if one neglects electron-electron interactions, the stability condition yields a finite width of the strip given by
\be a_k=\bigg(\frac{2\kappa \Delta E}{\pi^2 e^2 \frac{d n_{\rm el}(x)}{dx} |_{x_k}}\bigg)^{1/2}, \ee
where $\kappa$ is the dielectric constant of the material and $\Delta E$ is determined by the single particle energy gap. The gap can be either the Landau ($\hbar \omega_c$) or Zeeman ($g^*\mu_BB$) gap, where $\omega_c=eB/m$ is the cyclotron frequency and $\mu_B$ is the Bohr magneton together with the effective $g^*$ factor. We specify the strength of the gap by $\alpha=g^*\mu_BB/\hbar \omega_c$, if the gap is due to Zeeman splitting; otherwise is given by 1-$\alpha$. Note that, at Si/SiGe hetero-structures an additional gap exists due to valley degeneracy. Although, the above non-self-consistent scheme seems to be reasonable in handling the electrostatics of the 2DES, numerical self-consistent calculations show that the Chklovskii picture fails to describe the electron distribution~\cite{Oh97:13519,Afif:aysm}. This is due to the oversimplified assumptions of boundary conditions and essentially is due to the fact that the 2DES is by no means a perfect metal, as considered at the mentioned work. Even utilizing the above boundary conditions, self-consistency alters the estimated positions and the widths strongly~\cite{siddiki2004}. We will re-present the self-consistency below. However, with a slight modification of the density distribution inspired by the self-consistent calculations one can still obtain the widths that coincide with the experimental findings~\cite{afifABI:10}. We describe the electron density by
\be n_{\rm el}(x)=(1-e^{-(x-l_d)/t})n_0, \ee
where $t$ determines the width of the electron poor region in front of the metallic contacts. The widths can be obtained as
\be a_k=\sqrt{\frac{4 a_B^*\alpha}{\pi \nu_{0}}\frac{t}{e^{-(x_k-l_d)/t}}}, \label{eq:ak}\ee
$a_B^*$ being the effective Bohr radius.
\begin{figure}[t!]
 {\centering
 \includegraphics[width=.8\linewidth]{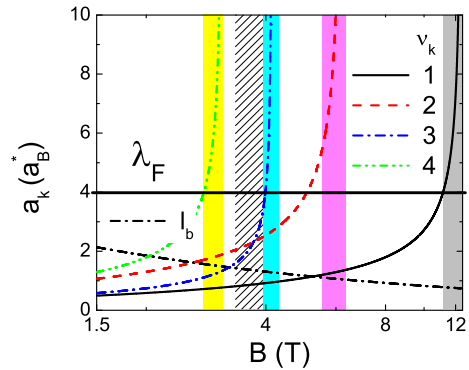}
 \caption{ \label{fig:fig1} Analytically calculated incompressible widths in units of $a_B^*$ ($\approx9.81$ nm for GaAs). The Fermi wavelength $\lambda_F$ considering a typical density of $3\times10^{15}$ m$^{-2}$, horizontal line. The shaded boxes indicate the quantized Hall interval, whereas diagonally shaded region depicts the overshoot interval. We set $t=7~a_B^*$ and $l_d=20a_B^*$, leading similar results with the self-consistent calculations}}
\end{figure}
In Fig.~\ref{fig:fig1}, we show the calculated widths of the incompressible strips with $\nu=1..4$, considering a fixed width of the electron poor region and an effective $g^*$, enhanced by the exchange interactions. We observe that more than one incompressible strip with different $\nu$ can co-exist below $\nu=2$, which we will reconsider their evanescent properties later. For the moment, we would like to calculate the amount of current confined to these strips. Note that, within the incompressible strips the drift velocity ($v_d(x)\propto \textbf{E}\times \textbf{B}$) is finite, whereas at the compressible strips it is zero, since the Hall field $E_H(x)$ equals to zero, due to perfect screening. Then, the current density can be calculated via $j_y(x)=-en_{\rm el}(x)v_d(x)$. Recall that, the electron density is constant within the strip, whereas $v_d(x)$ is determined by the electrochemical potential difference between the source and drain contacts. Hence, one can write the total current carried by the strip $k$ as
\be I_k=\int_{-a_k/2}^{a_k/2} j_y^k(x)dx= \frac{e}{2\pi}\frac{k eB}{m} \ee
Now to determine the Hall voltage, one should also know the local conductivities. The longitudinal resistivity (or resistance if a square geometry is assumed) is given by
\begin{figure}[t]
 {\centering
 \includegraphics[width=.8\linewidth]{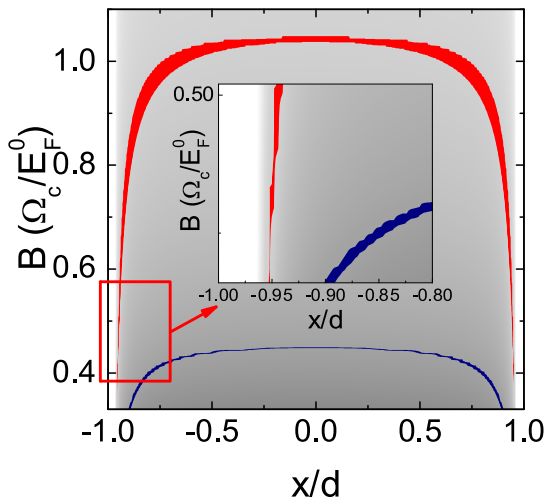}
 \caption{\label{fig:fig2}The evolution of the incompressible strips as a function of normalized lateral coordinate (horizontal axis) and $B$ field (vertical axis), considering $\nu=2$ dark (red) and $\nu=4$ darker (blue). We set $l_d/d=0.02$, where $d=10$ $\mu$m, at default temperature $\Theta_d=kT/E_F^0=0.01$. Inset depicts the overlap region.}}
\end{figure}
\be \rho_l(x)=\frac{1}{\bigg(\sigma_l(x)+\sigma_H^2(x)/\sigma_l(x)\bigg)},\ee
since the longitudinal conductivity $\sigma_l(x)$ goes to zero at the incompressible strip~\cite{Ando82:437} and Hall conductivity $\sigma_H(x)=\frac{e^2}{h}\nu(x)$, the only contribution to the Hall voltage comes from this region~\cite{siddiki2004}. The only unresolved problem is now to obtain the quantized Hall voltage, which is not the case if $k>2$ since it equals to
\be V_H=\frac{e^2}{h}\sum_k{I_k/k}, \ee
which is not quantized at all. Now we should reconsider the existence of more than one incompressible strip assuming different integer filling factors. In fact, having many incompressible strips is due to an artefact of the Thomas-Fermi approximation which fails to describe the electronic system at hand~\cite{Suzuki93:2986,siddiki2004}. If one considers the finite size of the wave-functions, together with the limitations of the Fermi distribution function at small systems, one immediately notices that if the incompressible strip becomes narrower than the Fermi wavelength it is no-longer a perfect channel without backscattering, \emph{i.e.} $\sigma_l\neq0$. Hence, only the incompressible strip having a width larger than the Fermi wavelength $\lambda_F$ can carry current. In Fig.~\ref{fig:fig1}, we also show the Fermi wavelength to distinguish the evanescent incompressible strips ($a_{k} <\lambda_F$) from the well developed ones. Now we can say that, if there exists an incompressible strip larger than $\lambda_F$, the Hall potential is quantized. We denote these intervals by shaded areas in Fig.~\ref{fig:fig1}. If $a_k$ becomes smaller than the magnetic length, then the system becomes fully classical and is simply described by the well known Drude formalism~\cite{afifABI:10}. The interesting transition takes place when the condition $l<a_k<\lambda_F$ is satisfied. In this situation, most of the current is flowing from the evanescent incompressible strip, with some background current spread all over the sample. In typical samples, due to the strong confinement at the edges or due to the strong disorder potential fluctuations, these evanescent incompressible strips immediately vanish, just after the quantized Hall plateau disappears. However, there might be cases where two or more of these evanescent strips survive and co-exist, satisfying $l_b<a_k<a_{k+1}...<\lambda_F$. Such a situation is also observed in Fig.~\ref{fig:fig1}, where $\nu=2$ and $\nu=3$ incompressible strips co-exist. For this case the Hall resistance will be given by
\be R_H=\frac{h}{e^2}(1/3+I_2/I(1/2+1/3)), \ee satisfying the current conservation, namely $I=I_2+I_3$. The second term in the bracket reflects the contribution to the current from the $\nu=2$ evanescent incompressible strip and one can easily see that, the Hall resistance is higher than the quantized value of $ \frac{h}{3e^2}$. This is exactly the case of the resistance overshoot: it occurs at the lower $B$ side of the plateau and strongly depends on the system parameters like the Fermi wavelength and geometry (\emph{i.e.} edge profile)~\cite{Shl2005,sailer:10}.

Having determined the conditions to observe overshoot depending on the existence and properties of the incompressible strips, next, we investigate overshoot within the framework of interaction theory of the quantized Hall effect.

\section{The self-consistent scheme}
In this section we summarize the numerical calculation algorithm that provides a consistent explanation to the IQHE. We consider a 2DES confined to the interval
$-d < x < d$, where $d$ is the half-width of the sample. The repulsive Coulomb interaction among the electrons is described by the Hartree potential,
\begin{equation}
V_{\rm H}(x)=\frac{2e^2}{\kappa}\int_{-d}^d dx'K(x,x')n_{\rm el}(x').
\label{eq:Vhatree}\end{equation}
Here $K(x,x')$ is the kernel satisfying the boundary
conditions, $V(-d)=V(d)=0$, given as
\begin{equation}
K(x,x^{'})=\ln\left|\frac{\sqrt{(d^2-x^2)(d^2-{d'^{2}})+d^2-x^{'}x}}{(x-x^2)d}\right|.
\end{equation}
Then the total potential (energy) of the electron is determined by
\begin{equation}
V(x)=V_{\rm bg}(x)+V_{\rm H}(x),
\end{equation}
where, the first term is the background potential describing the
external electrostatic confinement due to the donors and is given by
\begin{equation}
V_{\rm bg}(x)=-E_0\sqrt{1-(x/d)^2}.
\end{equation}
Here $E_0=2\pi e^2n_0d/\kappa$ is the minimum of the confinement.
 \begin{figure}[t!]
 {\centering
 \includegraphics[width=.8\linewidth]{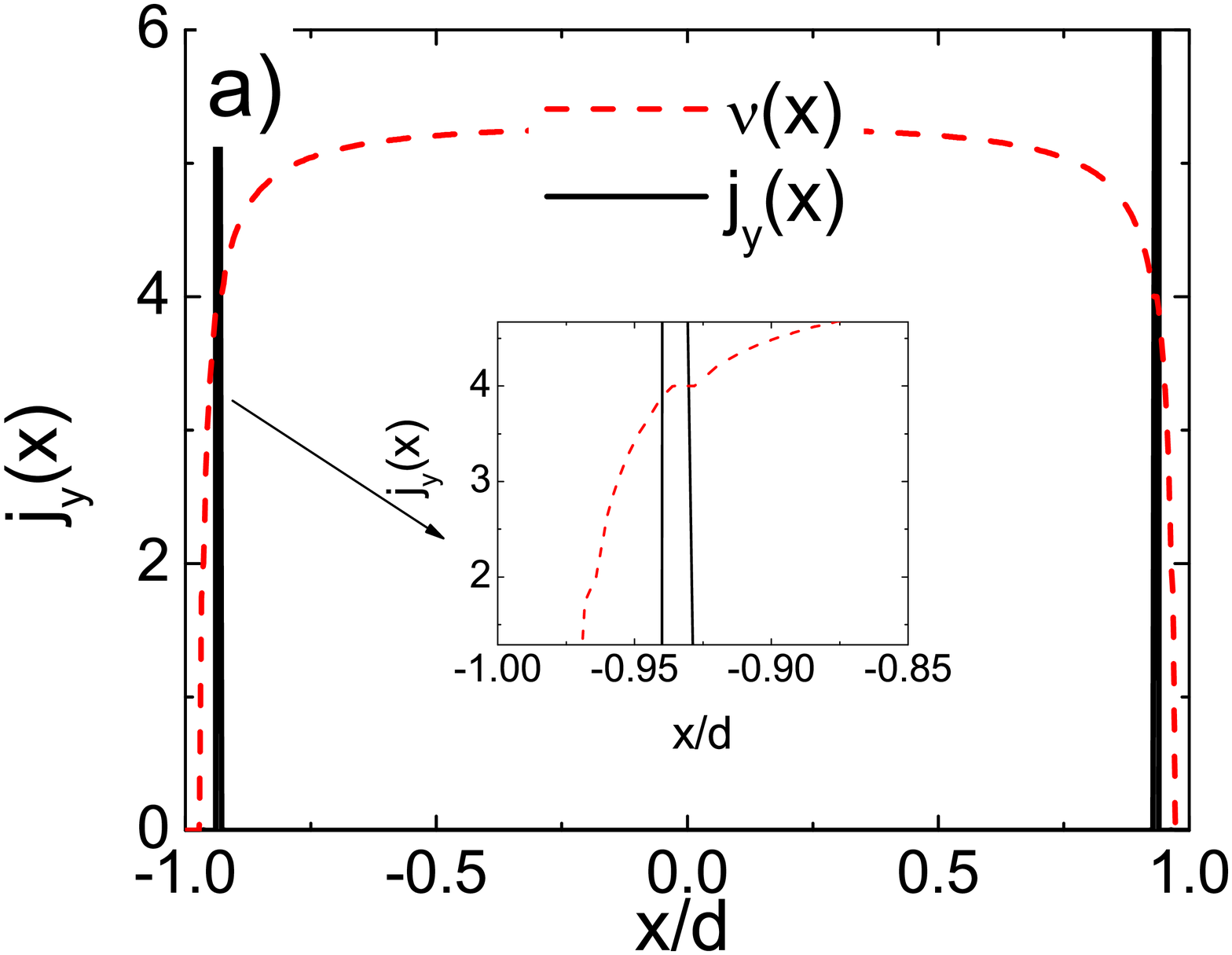}
  \includegraphics[width=.8\linewidth]{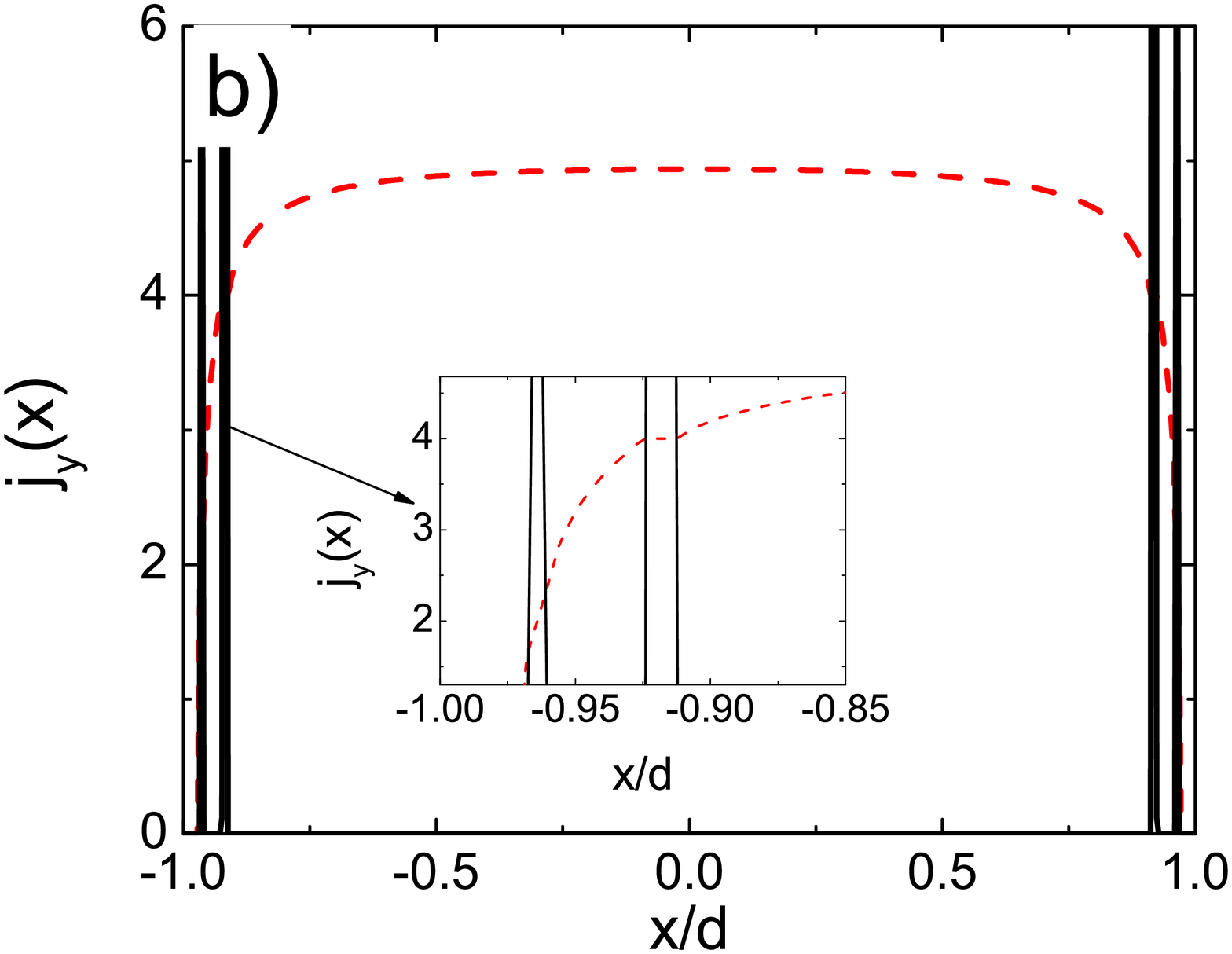}
   \includegraphics[width=.8\linewidth]{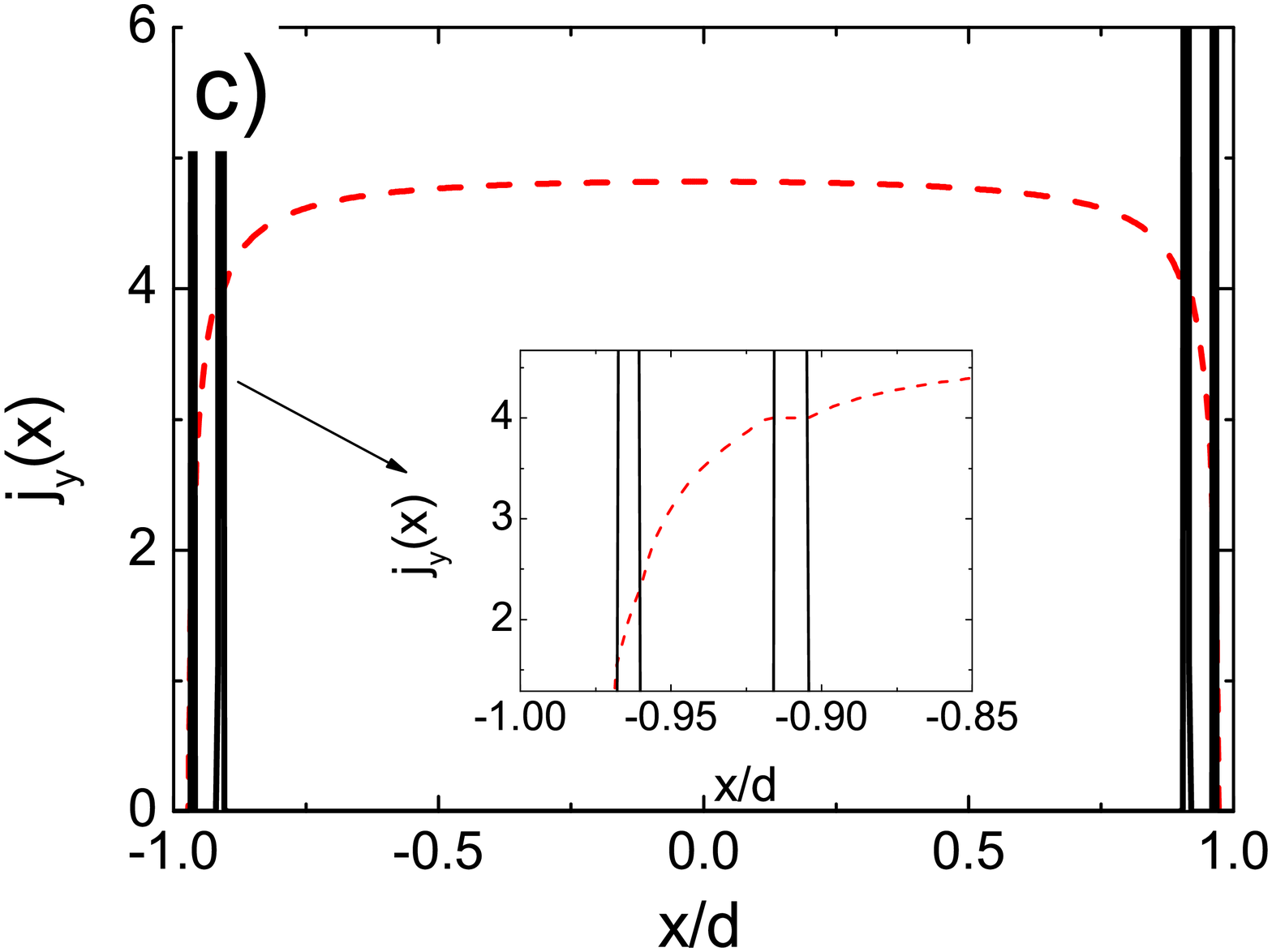}
 \caption{\label{fig:fig3}The current density (solid line) and filling factor distribution (broken line) as a function of lateral coordinate, calculated at (a) B=0.38, (b) 0.405 and (c) 4.15. Considering default temperature and same parameters in Fig.~\ref{fig:fig2}.}}
 \end{figure}
The solution involves the self-consistent determination of the electron density via
\begin{equation}
n_{\rm el}(x)=\int dED(E)f(E+V(x)-\mu^*)
\end{equation}
which is valid in the approximation of a slowly-varying potential, the namely Thomas-Fermi approximation (TFA). Here,
$f(E)=1/[\exp(E/k_{B}T)+1]$ is the Fermi distribution function with $k_{B}$
Boltzmann constant. The density of states $D(E)$ is to be taken from self-consistent Born approximation~\cite{Ando82:437} and $\mu^*$
is the constant equilibrium electrochemical potential. Since, the overshoot effect is independent of the actual origin of the single particle gap, from now on we assume spin degeneracy and neglect Zeeman splitting. The density of states (DOS) and local conductivities are determined assuming an impurity potential having a Gaussian form~\cite{Ando82:437}
\begin{equation}
V(r)=\frac{V_I}{\pi R^2}\exp(-\frac{r^2}{R^2})
\end{equation}
where the range $R$ is of the order of the spacing between
2DES and doping layer, together with the impurity strength $V_I$. In strong magnetic fields, the Landau levels are broadened due to the scattering from the impurities and the level width is given by
\begin{equation}
\Gamma^2=4\pi n_I^2 V_I^2/(2\pi l^2)=(2/\pi)\hbar\omega_c\hbar/\tau,
\end{equation}
where $n_I$ is the number density of the impurities and $\tau$ is the momentum relaxation time. We express the widths by the magnetic energy to
characterize the impurity strength by the dimensionless ratio
$\gamma =\Gamma /\hbar \omega_c$ and define the strength parameter as calculated at 10 T as
\begin{equation}
\gamma_I=[(2n_IV_0^2m^*/\pi \hbar^2)(1.73 ~{\rm meV})]^{1/2}.
\end{equation}
The above set of equations allow us to determine the electron density, electrostatic potential and local conductivities in a self-consistent manner when solved numerically by means of successive iterations. The details of the calculation scheme is described in detail elsewhere~\cite{siddiki2004}.

Fig.~\ref{fig:fig2} depicts the calculated filling factor distribution as a function of normalized spatial coordinate $x/d$ and external field $B$ expressed in units of $\Omega_c/E_F^0=\hbar\omega_c/E_F^0$, where $E_F^0$ is the Fermi energy at the center of the sample. The dark colored croissant like areas highlight the distribution of the incompressible strips. One can see that, two evanescent incompressible strips with different filling factors, namely $\nu=2$ and $4$, co-exist in the interval $0.4<B<0.48$. In the following, we will focus on such intervals to seek resistance overshoot.

Starting from the self-consistent quantities we calculate the current distribution within the local Ohm's law, relating the local electric field $\vec{E}(x,y)$ to the current density $\vec{j}(x,y)$ via $\vec{E}(x,y)=\hat{\sigma}(x,y)\vec{j}(x,y)$. A typical current distribution is shown in Fig.~\ref{fig:fig3} considering three characteristic $B$ field values, where a) most of the current is confined to the inner strip b) the current is shared by both strips, however, flows mainly from the inner strip and c) the current is almost equally confined to both strips. We would expect to have resistance overshoot in cases b) and c), whereas a) would present the classical Hall effect.

\begin{figure}[t]
 {\centering
 \includegraphics[width=.8\linewidth]{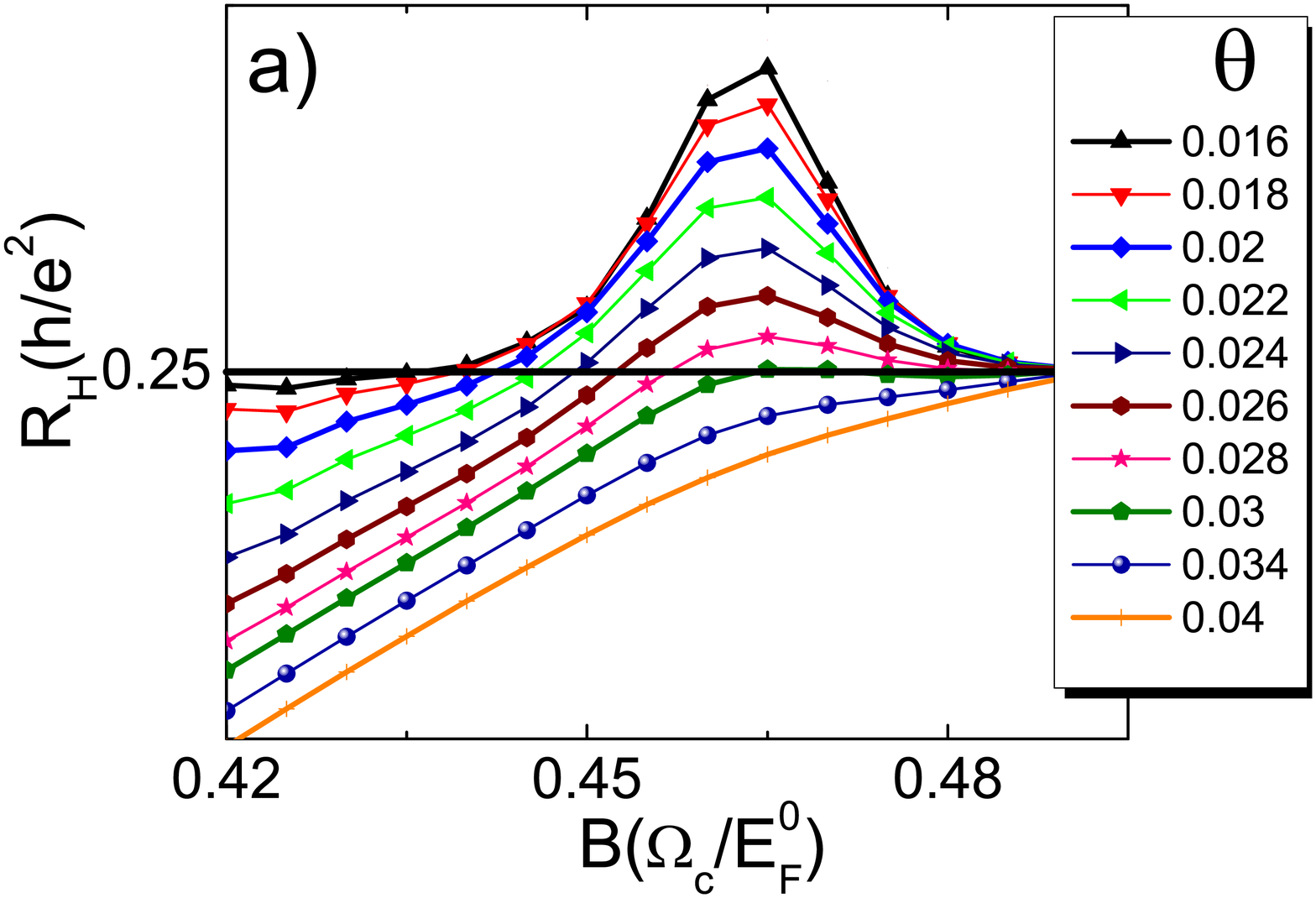}
  \includegraphics[width=.8\linewidth]{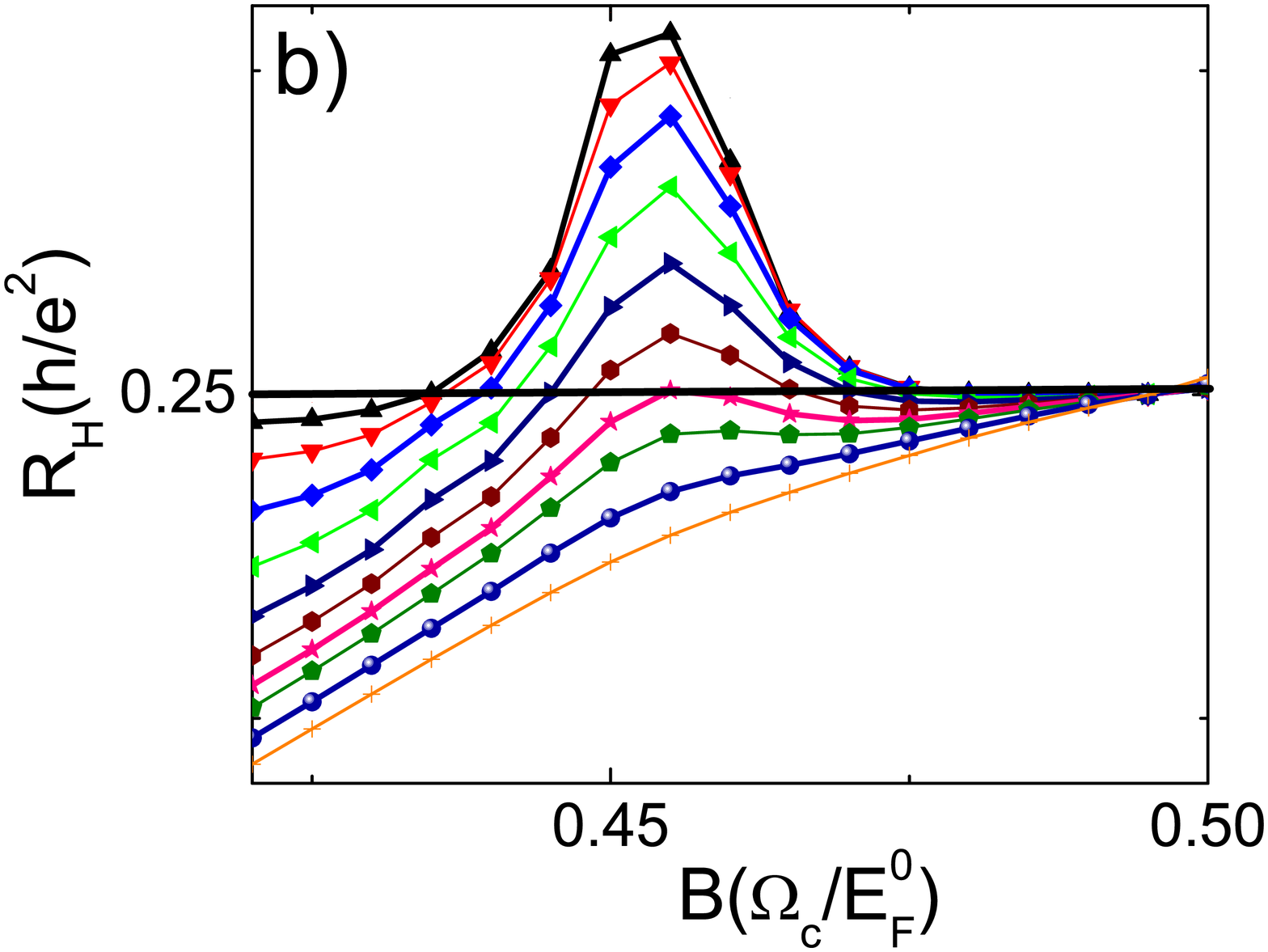}
 \caption{ \label{fig:fig4} Magnetic field dependence of the Hall resistance $R_H$ at different scaled temperatures $\Theta=k_{B}T/E_{F}^{0}$ with the sample width of (a) $2d=8$ $\mu $m and (b) $2d=12$ $\mu$m, $\gamma_{I}=0.3$.}}
 \end{figure}

We depict the calculated Hall resistance as a function of $B$ in Fig.~\ref{fig:fig4}, here we consider two different sample widths and calculations are performed at various temperatures. One can clearly see the overshoot at the expected $B$ intervals. The overshoot is smeared by the increase of the electron temperature, since the evanescent incompressible strip assuming $\nu=2$ is washed out due to the condition $a_2<l_b$. This finding also coincides with the experimental results showing that, the overshoot disappears with increasing temperature.

\begin{figure}[t]
 {\centering
 \includegraphics[width=.8\linewidth]{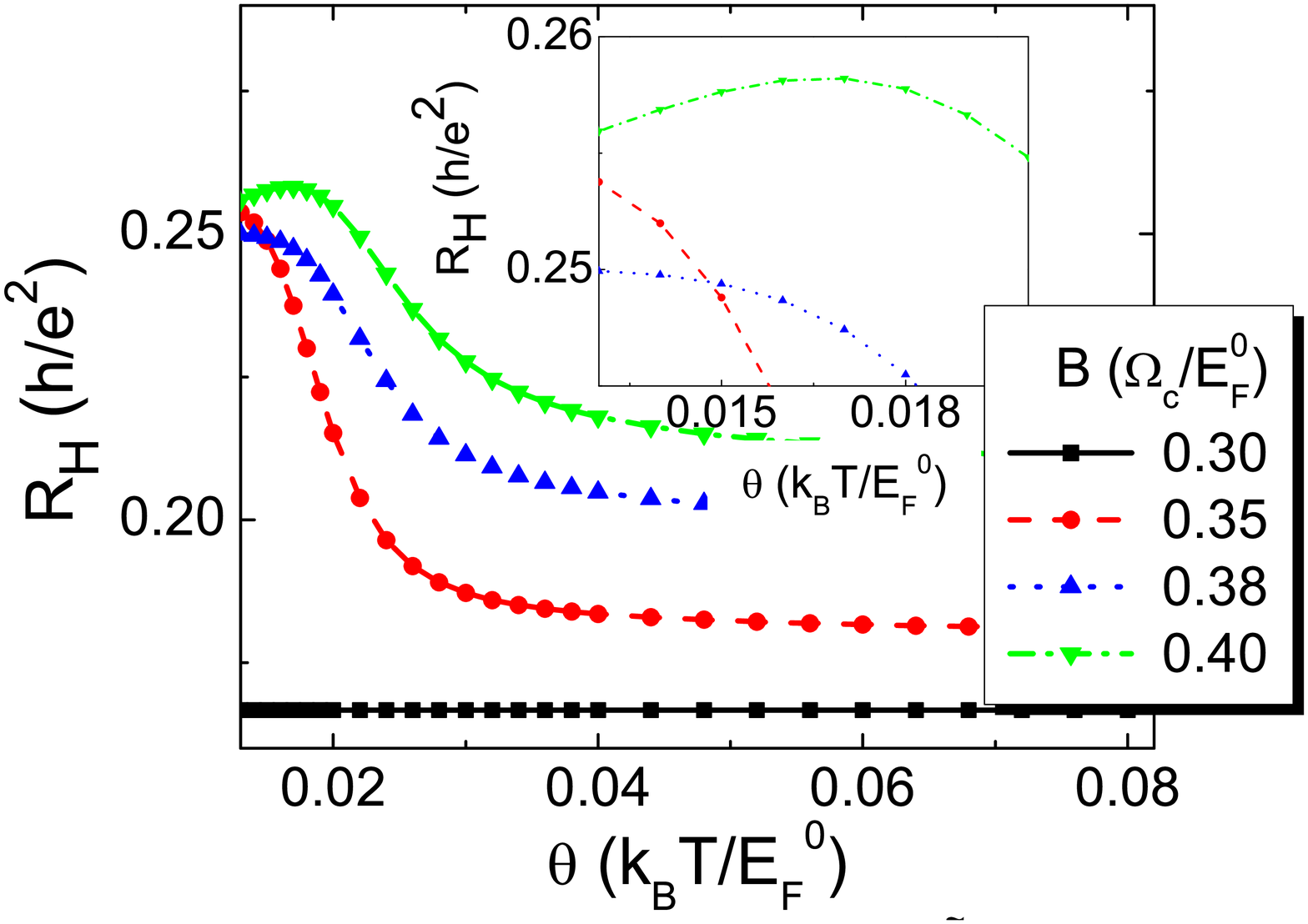}
 \caption{ \label{fig:fig5} Calculated Hall resistance $R_{H}$ versus scaled temperature ($\Theta$), for different values of magnetic field, ($\Omega_{c}/E_{F}^{0}$). The sample parameters are $2d=10$ $\mu m$ and $\gamma_{I}=0.05$.}}
 \end{figure}

\begin{figure}[t]
 {\centering
  \includegraphics[width=.8\linewidth]{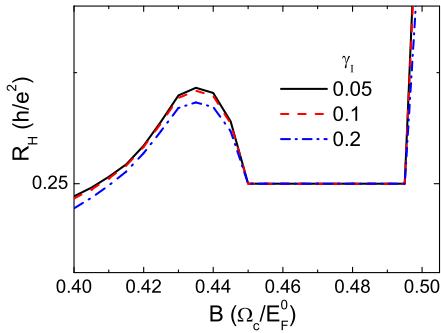}
 \caption{\label{fig:fig6}Hall resistance versus magnetic field, calculated for different level broadening $\gamma_I$ considering an 4 $\mu$m wide sample at default temperature and depletion length.}}
\end{figure}

We investigate the effect of temperature on the overshoot in a more detailed manner in Fig.~\ref{fig:fig5}, considering a 10 $\mu$m wide sample and for three characteristic values of the $B$ field. One sees that, the Hall resistance is impregnable to small temperature variations if the system is out of the overshoot regime (black solid line with boxes). Meanwhile at the overshoot interval $R_H$ depends strongly on the temperature, broken lines, which is even pronounced at the peak maximum.
We finalize our discussion, by presenting the effects of the impurity strength on the overshoot in Fig.~\ref{fig:fig6}. Remarkably, the disorder has a minor influence on the amplitude of the overshoot, which is seen by the weak dependence of the peak depending on $\gamma_I$. This is mainly due to the fact that, the impurities only shrink the widths of the evanescent incompressible strips via level broadening. However, it's effect negligible when compared to the role of temperature. It is important to note that, we did not include the long-range potential variations to our screening calculations, which are known to be influential in determining the position and stability of the quantized Hall plateaus~\cite{Siddiki:ijmp}. The effects resulting from long-range fluctuations is an open question, which we would like to attack in the near future.
To summarize, we have shown that the resistance overshoot can be obtained within the framework of self-consistent screening theory. This calculation scheme allows us to investigate the effects of various experimental parameters on the overshoot. It is observed that, the overshoot depends strongly on temperature, in contrast, is immune to short-range impurity scattering. From our sample width dependent calculations, we conclude that if the edge effects are dominant the overshoot is enhanced. Explicitly, for the large samples disorder effects become more important and overshoot tends to disappear.
\section{Predictions and Conclusions}
In the light of above results and discussions we predict that, for the smooth edge defined samples the overshoot effect should be enhanced. The reason is: To have co-existing evanescent incompressible strips the condition $l_b<a_k,a_{k+1}<\lambda_F$ should be satisfied, this can only happen if the electron density varies slowly, so that the derivative in Eq.~\ref{eq:ak} becomes small. Hence the strip becomes large. The experimental test can be as follows, one can define two narrow (\emph{e.g.} $2d\sim10~\mu$m) Hall bars residing parallel to each other, where one of the Hall bars is defined by shallow etching and the other by deep etching. Since, in principle, all the intrinsic properties of the material would be the same for both samples the observed difference at the overshoots (enhanced at the shallow sample) would point out the effects due to the formation of wide evanescent incompressible strips. A gate defined sample can be utilized as well, similar to the ones reported in the literature~\cite{EPL:exp,Hor2008}.

Another interesting observation would be generating overlap of fractional-integer or fractional-fractional evanescent incompressible strips. For the overlap of $\nu=1/3$ and $\nu=1$ one would observe a resistance overshoot at the end of filling factor one plateau. However, to keep $\nu=1/3$ strip at the edge evanescent is fairly difficult, due to small many-body gap, for large magnetic field intervals. On the other hand, by invoking metallic gates parallel to the edges and biasing them positively will result in an enhancement of the fractional strip widths. We claim that, one can observe an overshoot at the lower end of the $\nu=1$ or $\nu=2/3$ plateau due to an evanescent incompressible strip of $\nu=1/3$ supported by the reconstruction of the edge-electrostatics due to parallel metallic strips. For sure, such samples require fairly high mobilities ($> 2.0\times10^6$ V/cm.s).

In summary, we analyzed the temperature, disorder, magnetic field and sample width effects on the overshoot resistance considering the GaAs/GaAlAs heterojunction.  We obtained that with increasing both the temperature and disorder strength the overshoot peak decreases. The calculations show that the overshoot resistance as a function of magnetic field depends strongly on the edge electrostatics of the sample. Observation of enhanced resistance overshoot considering integer and fractional states is predicted by manipulating the edge potential profile.
\acknowledgments
We highly appreciate M. Grayson for introducing us the overshoot problem and for fruitful discussions. D. Bourgard and J. Sailer is acknowledged for giving us the detailed experimental insight, in particular A. Wild for initiating the semi-classical model and his critical questions. This work is supported by the scientific and technological research council of Turkey under grant (TBAG:109T083) and Istanbul university IU-BAP:6970.
\bibliographystyle{eplbib}
%\bibliography{zitate}

%\bibliographystyle{elsart-num}
%\bibliography{cite}
%\bibliography{siddiki}

\end{document}